\documentclass[iop,apjl,numberedappendix]{emulateapj-rtx4}
\usepackage{graphicx}
\usepackage{mathrsfs}
\usepackage[intlimits,centertags]{amsmath}
\usepackage{amssymb,amsfonts}
\usepackage[pdftex]{hyperref}
\usepackage[x11names]{xcolor}

\hypersetup{pdftitle={Small-Scale Anisotropies of Cosmic Rays from Relative Diffusion},
pdfsubject={Small-Scale Anisotropies of Cosmic Rays from Relative Diffusion},
pdfauthor={Markus Ahlers and Philipp Mertsch},
pdfstartview={FitH},
colorlinks=true,
bookmarksopen=false,
bookmarksnumbered=false,
bookmarksopenlevel=0,
linkcolor=Blue1!70!black,
citecolor=Blue1!70!black,
urlcolor=Blue1!70!black
}

\begin{document}

\title{Small-Scale Anisotropies of Cosmic Rays from Relative Diffusion}
\author{Markus Ahlers}
\affiliation{WIPAC \& Department of Physics, University of Wisconsin--Madison, Madison, WI 53706, USA\vspace{-0.2cm}
}
\author{Philipp Mertsch}
\affiliation{Kavli Institute for Particle Astrophysics \& Cosmology, Menlo Park, CA 94025, USA}

\begin{abstract}
The arrival directions of multi-TeV cosmic rays show significant anisotropies at small angular scales. It has been argued that this small-scale structure can naturally arise from cosmic ray scattering in {\it local} turbulent magnetic fields that distort a global dipole anisotropy set by diffusion. We study this effect in terms of the power spectrum of cosmic ray arrival directions and show that the strength of small-scale anisotropies is related to properties of relative diffusion. We provide a formalism for how these power spectra can be inferred from simulations and motivate a simple analytic extension of the ensemble-averaged diffusion equation that can account for the effect.
\end{abstract}


\maketitle

\section{Introduction}

The propagation of cosmic rays (CRs) through turbulent magnetic fields is effectively described as a diffusion process. This approach provides an excellent description of the observed spectra and chemical abundances of Galactic CRs up the CR knee at a few PeV. The arrival direction of CRs is mostly isotropized in this process and the location of the sources is hidden. The first order correction to the isotropic distribution is a small dipole anisotropy that scales with the local CR gradient according to Fick's law. The phase and strength of this dipole is expected to be a combined effect of the relative motion of the solar system with respect to the frame where CRs are isotropic~\citep{Compton:1935}, the density gradient of CRs, the strength of diffusion, and the strength and orientation of regular magnetic fields~\citep{Erlykin:2006ri,Blasi:2011fm,Pohl:2012xs,Mertsch:2014cua,Schwadron2014}.

Various CR observatories have revealed anisotropies in the arrival directions of TeV--PeV Galactic CRs on large and small angular scales~\citep{Amenomori:2005dy,Amenomori:2006bx,Guillian:2005wp,Abdo:2008kr,Abdo:2008aw,ARGO-YBJ:2013gya,Aglietta:2009mu,Abbasi:2011ai,Aartsen:2013lla,Abeysekara:2014sna}. In this Letter we will focus on small-scale anisotropies that have been studied via a power spectrum analysis by IceCube~\citep{Aartsen:2013lla} and HAWC~\citep{Abeysekara:2014sna} (see Fig.~\ref{fig2}). \cite{Giacinti:2011mz} have argued that small-scale anisotropies can naturally emerge as a next-to-leading order correction of the diffusive approximation if one takes into account the local scattering of CRs within the diffusive scattering length. \cite{Ahlers:2013ima} showed that under certain simplifying assumptions these small-scale fluctuations necessarily arise from the application of Liouville's theorem. Assuming a simple hierarchical evolution equation of the multipoles predicts a power spectrum that matches well the observations.

Fluctuations of the arrival direction of CRs induced by local turbulent fields are not directly accessible in standard diffusion theory. This becomes obvious if one considers the angular power spectrum of the CR phase-space distribution (PSD) $f(t,{\bf r},{\bf p})$ defined as
 \begin{equation}\label{eq:Celldef}
C_\ell = \frac{1}{4\pi}\int {\rm d}\hat{\bf p}_1 \int {\rm d}\hat{\bf p}_2P_\ell(\hat{\bf p}_1  \hat{\bf p}_2) f_1f_2\,,
\end{equation}
where $\hat{\bf p}_{i}$ denotes a unit vector of ${\bf p}_{i}$ and $P_\ell$ are Legendre polynomials of degree $\ell$. Here and in the following we use the abbreviation $f_i = f(0,{\bf 0},{\bf p}_i)$ for the local PSD. In general, the small-scale magnetic field that determines the PSD is unknown. Therefore, the PSD can only be predicted as an average over a statistical ensemble of random small-scale magnetic fields, usually characterized by its power spectrum. In ergodic systems, the average over turbulent magnetic field configurations, {\it i.e.},~the {\it ensemble} average denoted in the following by $\langle \mathellipsis \rangle$, is equivalent to a time-average. However, the ensemble average of $\langle C_\ell\rangle$ is proportional to $\langle f_1 f_2\rangle$, which is {\it larger} than the product of the ensemble averages, {\it i.e.},~the standard solution of diffusion theory.

\section{Relative Diffusion}
\label{sec:reldiff}

On large timescales, the motion of CRs in turbulent magnetic fields can be effectively approximated as a diffusion process. The angular-integrated PSD $n = n(t,{\bf r}) \equiv \int {\rm d}\hat{\bf p} f(t,{\bf r},{\bf p})$ then satisfies a diffusion equation, $\partial_t n \simeq \nabla({\bf K}\nabla n)$, where the diffusion tensor ${\bf K}$ can be expressed as
\begin{equation}\label{eq:K}
{K}_{ij} = \frac{\hat{B}_i\hat{B}_j}{3\nu_\parallel}+\frac{\delta_{ij}-\hat{B}_i\hat{B}_j}{3\nu_\perp}+\frac{\epsilon_{ijk}\hat{B}_k}{3\nu_A}\,.
\end{equation} 
Here, $\hat{\bf B}$ denotes a unit vector along the regular magnetic field. In general, the (scattering) rates $\nu_\parallel$, $\nu_\perp$ and $\nu_A$ are independent of one another. The presence of a small local gradient $\nabla{n}$ induces a dipole anisotropy in the quasi-stationary solution as a result of Fick's law, $4\pi \langle f\rangle \simeq n -3\hat{\bf p}{\bf K}\nabla n$. Therefore, standard diffusion theory predicts that the dipole strength of the relative intensity $\delta I\equiv(4\pi f-n)/n$ is given as
\begin{equation}\label{eq:standardC1}
\frac{1}{4\pi}{C_1} = \left|\frac{{\bf K}\nabla n}{n}\right|^2\,.
\end{equation}
We will see in the following that this relation becomes modified once we consider corrections of the PSD product in the ensemble average. This will also introduce multipoles at small angular scales, which can be related to properties of relative diffusion.

We now study the effect of small local fluctuations of the PSD around the ensemble average, $\delta{f} = f- \langle f\rangle$. According to Liouville's theorem we can relate the local ({\it i.e.},~${\bf r}=0$) PSD $f_i = f(0,{\bf 0},{\bf p}_i)$ to the contribution backtracked along CR trajectories to an arbitrary time, 
\begin{multline}\label{eq:backtrack}
4\pi f_i \simeq 4\pi\delta{f}(-T,{\bf r}_i(-T),{\bf p}_i(-T)) \\ + n 
+ [{\bf r}_i(-T)-3\hat{\bf p}_i(-T){\bf K}]\nabla n\,,
\end{multline}
where $n$ and $\nabla n$ denote the local CR density and gradient and ${\bf r}_i(-T)$ and ${\bf p}_i(-T)$ are the position and momentum of a CR (that is at position ${\bf r}_i = {\bf 0}$ and $\hat{\bf p}_i(0)=\hat{\bf p}_i$ at time $t = 0$) at backtracking time $T$. Now, in the limit of large $T$ the last term in Eq.~(\ref{eq:backtrack}) is dominated by the third term scaling with the position of the particle. Also, for two momenta ${\bf p}_1\neq{\bf p}_2$ we can assume that the ensemble average of fluctuations are uncorrelated, $\langle\delta{f}_1(-T)\delta{f}_2(-T)\rangle \simeq 0$, for sufficiently large backtracking times when the CR trajectories eventually separate. 
In the degenerate case ${\bf p}_1={\bf p}_2$ the two backtracked CR trajectories stay correlated over arbitrarily long backtracking times. It will be sufficient to assume that $\langle (\delta f(-T))^2\rangle$ remains finite. We can then express the multipole spectrum of the ensemble-averaged relative intensity as the limit
\begin{multline}\label{eq:Cell}
\frac{1}{4\pi}{\langle C_\ell\rangle}\simeq \int \frac{{\rm d}\hat{\bf p}_1}{4\pi} \int \frac{{\rm d}\hat{\bf p}_2}{4\pi}P_\ell(\hat{\bf p}_1  \hat{\bf p}_2)\\\times \lim_{T\to\infty}\langle {r}_{1i}(-T){r}_{2j}(-T)\rangle \frac{\partial_i n\partial_j n}{n^2}\,,
\end{multline}
$\partial_i$ being shorthand for $\partial/\partial x_i$.

Note that the $\ell\geq1$ multipole spectrum is generated through {\it relative} diffusion: it can be easily seen that the sum over {\it all} ensemble-averaged multipoles of the relative intensity
can be expressed via the symmetric part of the diffusion tensor $\langle{r}_{i}(-T){r}_{j}(-T)\rangle \to 2T{K}^{\rm s}_{ij}$ in the limit of large backtracking times $T$,
\begin{equation}\label{eq:id1}
\frac{1}{4\pi}\sum_{\ell\geq0}(2\ell+1){\langle C_\ell\rangle}(T) \simeq 2T{K}^{\rm s}_{ij}\frac{\partial_i n\partial_j n}{n^2}\,.
\end{equation}
On the other hand, the average monopole contribution in this limit can be expressed as 
\begin{equation}\label{eq:C0}
\frac{1}{4\pi}{\langle C_0\rangle}(T) \simeq 2T\left({K}^{\rm s}_{ij}-\widetilde{K}^{\rm s}_{ij}\right)\frac{\partial_i n\partial_j n}{n^2}\,,
\end{equation}
where the symmetric part of the relative diffusion tensor is defined as
\begin{multline}\label{eq:Krel}
\widetilde{K}^{\rm s}_{ij} =\int\frac{{\rm d}\hat{\bf p}_1}{4\pi}\int\frac{{\rm d}\hat{\bf p}_2}{4\pi}\\\times \lim_{T\to\infty} \frac{1}{4T}\big\langle\Delta{r}_{12i}(-T)\Delta{r}_{12j}(-T)\big\rangle\,,
\end{multline}
with $\Delta{\bf r}_{12} \equiv {\bf r}_1-{\bf r}_2$. Therefore, the sum of multipoles $\ell\geq1$ is related to the relative diffusion tensor. For uncorrelated particle trajectories, this expression reduces to the normal diffusion tensor. However, particle trajectories with a small relative opening angle will follow similar trajectories and the relative contribution (\ref{eq:Krel}) remains small over long timescales. Note that the multipoles in Eq.~(\ref{eq:Cell}) are expected to be finite in the limit of large backtracking times since particle trajectories with arbitrarily small opening angles will eventually become uncorrelated, $\langle {r}_{1i}(-T){r}_{2j}(-T)\rangle \to 0$. 

\begin{figure}[t]\centering
\includegraphics[width=\linewidth]{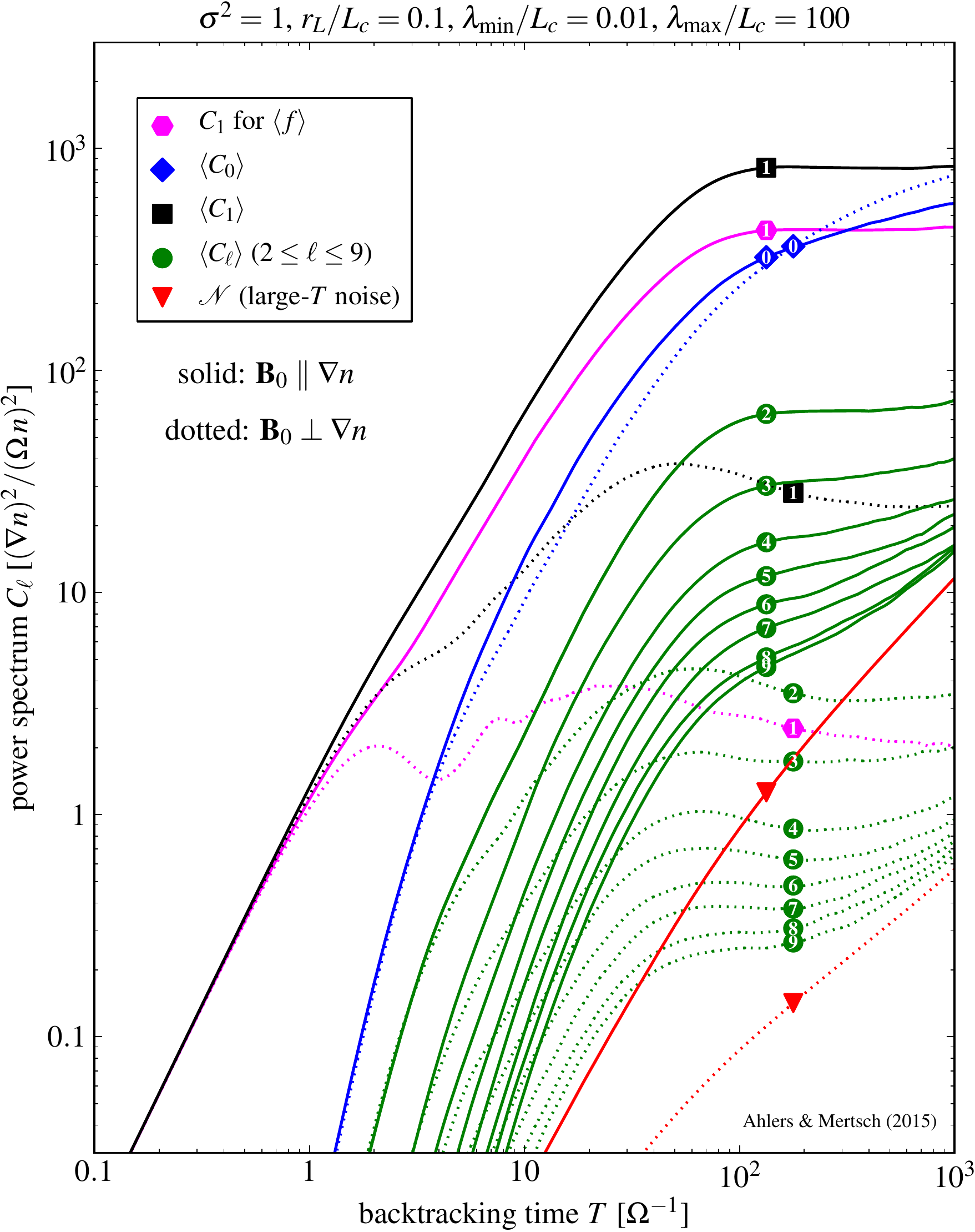}
\caption[]{Evolution of the ensemble-averaged power spectrum (\ref{eq:Cell}) for a CR gradient parallel (solid lines) and perpendicular (dotted lines) to the regular magnetic field and the 3D turbulence model discussed in the main text. We show results in terms of the dipole $\langle C_1\rangle$ (black), monopole $\langle C_0\rangle$ (blue), and medium-$\ell$ multipoles (green). We also show the asymptotic noise level (\ref{eq:noise}) (red) and the dipole prediction (\ref{eq:standardC1}) of standard diffusion (magenta) evaluated by the replacement $\langle{r}_{1i}{r}_{2j}\rangle \to \langle{r}_{1i}\rangle\langle{r}_{2j}\rangle$ in Eq.~(\ref{eq:Cell}).}\label{fig1}
\end{figure}

\section{Simulation}
\label{sec:simulations}

In the following, we will study the development of small-scale anisotropies via numerical simulations (see also \cite{Giacinti:2011mz,Ahlers2015,Lopez2015,Rettig2015}). We follow the approach of \cite{Giacalone:1999} and define a three-dimensional (3D) turbulent magnetic field as the sum $\delta{\bf B}({\bf r}) = \sum_{n=1}^N \delta {\bf B}_n \cos({\bf k}_n{\bf r}+\beta_n)$ with $N$ random phases $\beta_n$ and wave vectors ${\bf k}_n$ with 3D random orientations, on top of a regular field ${\bf B}_0$. The wave vector amplitudes $k_n$ range from $k_{\rm min}$ to $k_{\rm max}$ with equal logarithmic steps. The vectors $\delta {\bf B}_n$ have 3D random orientations subject to the conditions $\delta {\bf B}_n\perp{\bf k}_n$ and $|\delta {\bf B}_n|^2 \propto  k_n^3/(1+(k_nL_c)^\gamma)$ with a coherence scale $L_c$. We assume a Kolmogorov-type phenomenology with $\gamma = 11/3$ and the strength of the turbulence relative to the regular magnetic field is fixed via a parameter $\sigma^2$ and the normalization of the spatial average $\overline{\delta {\bf B}^2} = \sigma^2{\bf B}_0^2$.  

For the Galactic environment we expect a regular magnetic fields of the order of $6\,\mu$G and a turbulent component with $\sigma^2\lesssim1$ and a coherence scale $L_c\simeq100$~pc. The Larmor radius of 1-10~TeV CRs is then of the order of $r_L \simeq 10^{-5}L_c$ corresponding to a few hundred AU. A numerical study of this configuration with the methods described above is very time-consuming since the asymptotic behavior relevant for the multipole configuration is reached at a very late stage. Instead, here we will study a simple turbulent configuration that allows us to reach the necessary statistics to study the predicted effect qualitatively. We choose $\sigma^2=1$ with $\lambda_{\rm min}=0.01L_c$ and $\lambda_{\rm max}=100L_c$ with $k=2\pi/\lambda$ and fix the rigidity of the CRs to $r_L = 0.1L_c$. We sample over $120$ different turbulent magnetic field configurations. For each field configuration we uniformly sample $12288$ CR orientations $\hat{\bf p}_i(0)$ following the {\it HEALPix} parametrization ($n_{\rm side}=32$)~\citep{Gorski:2004by}, which we backtrack through the static magnetic field ${\bf B}({\bf r}) = B_0\hat{\bf e}_z+ \delta{\bf B}({\bf r})$ to find the initial position ${\bf r}_i(-T)$. 

Figure~\ref{fig1} shows the power spectrum at different backtracking times determined via a multipole expansion of the relative intensity sky maps using the {\it HEALPix} utilities. We show the cases of CR gradients parallel (solid lines) and perpendicular (dotted lines) to ${\bf B}_0$. The black lines show the evolution of the ensemble-averaged dipole induced by the CR gradient. The magenta lines correspond to the standard dipole anisotropy defined via Eq.~(\ref{eq:standardC1}) evaluated by the replacement $\langle{r}_{1i}{r}_{2j}\rangle \to \langle{r}_{1i}\rangle\langle{r}_{2j}\rangle$ in Eq.~(\ref{eq:Cell}). As expected, this dipole estimate is smaller than the ensemble average. Note that for the case of perpendicular diffusion (dotted lines) the difference in the asymptotic values is about one order of magnitude. The green lines show the multipole power for $2\leq\ell\leq9$. We also show the expected monopole component (\ref{eq:C0}) as blue lines. At large times, the high-$\ell$ components in the multipole expansion are dominated by noise, due to the finite number of trajectories. The noise level can be estimated for large backtracking times $T$ as pixel shot noise
\begin{equation}\label{eq:noise}
\mathcal{N}
 \simeq \frac{4\pi}{N_{\rm pix}}2TK^{\rm s}_{ij}\frac{\partial_in\partial_jn}{n^2}\,.
\end{equation}
This is indicated as red lines in Fig.~\ref{fig1} and clearly influences the level of high-$\ell$ multipoles (green lines) in the map. 

\begin{figure}[t]\centering
\includegraphics[width=\linewidth]{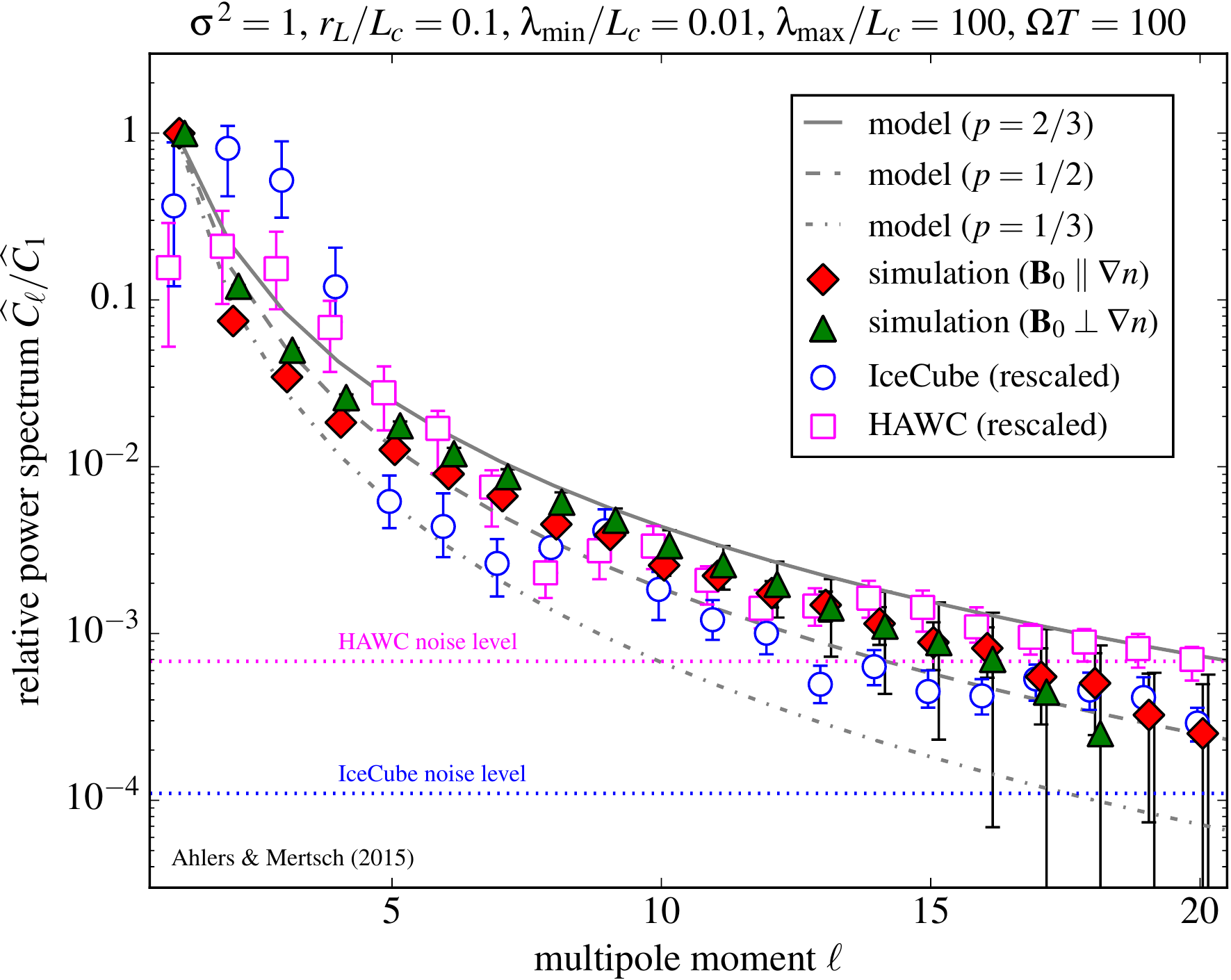}
\caption[]{Power spectrum estimator $\widehat{C}_\ell = \langle C_\ell\rangle -\mathcal{N}$ (normalized to $\widehat{C}_1$) for the parallel (filled red diamonds) and perpendicular (filled green triangles) CR gradient for the same simulations as shown in Fig.~\ref{fig1} at a backtracking time $\Omega T =100$. We also show the IceCube~\citep{Aartsen:2013lla} (blue open circles) and the HAWC~\citep{Abeysekara:2014sna} (magenta open squares) power spectra renormalized by a factor of $4\times10^6$ for better comparison. The different noise level of the data is indicated as dotted lines. The three gray lines correspond to the prediction of a relative scattering term $\nu_r(x) \propto (1-x)^p$ in Eq.~(\ref{eq:finalCl}) for three different values of $p$.}\label{fig2}
\end{figure}

The best estimator for the true power spectrum is then $\widehat{C}_\ell = \langle C_\ell\rangle-\mathcal{N}$ and the variance (excluding cosmic variance) is given by ${\rm var}(\widehat{C}_\ell) \simeq 2\mathcal{N}^2/(2\ell+1)$; see,{\it e.g.},~\cite{Campbell:2014mpa}. In Fig.~\ref{fig2} we show the estimators $\widehat{C}_\ell$ normalized to the dipole $\widehat{C}_1$ at a backtracking time $\Omega T=100$, corresponding to the timescale of the transition into the diffusion regime (cf.~Fig.~\ref{fig1}). Given the finite number of trajectories, it might be difficult to disentangle noise variance and cosmic variance in the simulations; we hence limit ourselves to the former only. Similar to the simulation, the sensitivity of experimental data to high-$\ell$ multipoles is limited to shot noise indicated as dotted lines in the plot. Overall, the simulation agrees well with the IceCube and HAWC data. For an isotropic Gaussian field, deviations at multipoles of $\ell \lesssim 5$ can be due to cosmic variance with ${\rm var}(\widehat{C}_\ell) \simeq 2 \widehat{C}_\ell^2/ (2 \ell + 1)$. Furthermore, the limited field of view of the observatories introduces additional systematic errors, in particular for low-$\ell$ multipoles, which are not included in the error bars of Fig.~\ref{fig2}.

\section{Generalized BGK Formalism}
\label{sec:BGK}

In the following, we derive the asymptotic form of the power spectrum via a microscopic description of relative diffusion. As before, we split the PSD into the average and fluctuating parts, $f = \langle f\rangle +\delta{f}$. The ensemble-averaged Boltzmann equation can be written as~\citep{JonesApJ1990}
\begin{equation}\label{eq:genBoltzmann}
\partial_t\langle f\rangle  + \hat{\bf p} \nabla\langle f\rangle  -i\boldsymbol{\Omega}{\bf L}\langle f\rangle  = \left\langle i\boldsymbol{\omega}{\bf L}\delta{f}\right\rangle\,.
\end{equation}
Here we have introduced the angular momentum operator ${L}_a = -i\epsilon_{abc}{p}_b\partial/\partial p_c$ and the rotation vectors ${\boldsymbol \Omega} = e{\bf B}_0/p_0$ and ${\boldsymbol \omega} = e\delta{\bf B}/p_0$.

The rhs of Eq.~(\ref{eq:genBoltzmann}) encodes the particle scattering in the random fields. The influence of the turbulence is typically approximated as a friction term introduced by \cite{Bhatnagar:1954zz} (BGK). In its original version this term drives the ensemble-averaged distribution $\langle f\rangle$ to its isotropic mean $n$ with an effective relaxation rate $\nu$, {\it i.e.},~$\left\langle i\boldsymbol{\omega}{\bf L}\delta{f}\right\rangle \simeq -\nu\left(\langle f\rangle-n/(4\pi)\right)$. The parameters $\nu_\parallel$, $\nu_\perp$, and $\nu_A$ are the effective parallel, perpendicular, and axial scattering rates, and can be related to the BGK rate $\nu$ via $\nu_\parallel = \nu$, $\nu_\perp/\nu_\parallel = 1+\Omega^2/\nu_\parallel^2$, and $\nu_A/\Omega = 1+\nu_\parallel^2/\Omega^2$. Substituting into Eq.~(\ref{eq:K}) we find that the standard $C_1^\parallel$ and $C_1^\perp$ (\ref{eq:standardC1}) scale as $1/{\nu_\parallel}$ and $1/{\nu_\perp}$, respectively, similar to the sum (\ref{eq:id1}) in the relative diffusion approach.

We can generalize the BGK approximation of the friction term to higher multipoles if we consider the effect of turbulence on the particle momenta as a Brownian motion on a sphere~\citep{Yosida1949}. For a distribution $f$ on a sphere the Brownian diffusion of momenta is described via $\partial_t f = -(\nu/2){\bf L}^2f$. In particular, if initially the configuration is localized, $f(0,{\bf p}) = \delta(\hat{\bf p}  \hat{\bf p}_0-1)$, then the solution at later times is given as $f(t,{\bf p}) = \sum_{\ell m}Y^*_{\ell m}(\hat{\bf p})Y_{\ell m}(\hat{\bf p}_0)e^{-\ell(\ell+1)\nu t/2}$. This reduces to a two-dimensional Gaussian distribution of the opening angle $\psi\ll1$ defined via $\cos\psi = \hat{\bf p}\hat{\bf p}_0$ with a width $\sigma^2 \simeq \nu t$. Therefore, for the rhs of Eq.~(\ref{eq:genBoltzmann}) we make the ansatz
\begin{equation}\label{eq:BGK}
\left\langle i\boldsymbol{\omega}{\bf L}f\right\rangle \simeq -(\nu/2){{\bf L}^2}\langle f\rangle,
\end{equation}
which is a natural generalization of the BGK approach. Note that this simple treatment of the scattering term is only approximate. Depending on the strength and coherence length of the turbulent magnetic field, the effective relaxation can in general be anisotropic even for a fully isotropic turbulence. However, in our case we are interested in the qualitative behavior and development of high-$\ell$ multipoles, and the first order BGK approximation seems appropriate.

Now, for the asymptotic solution of $\langle C_\ell\rangle$ at large backtracking times we have to look for stationary solutions of the equation
\begin{equation}\label{eq:CLevolution}
\partial_t\langle f_1f_2\rangle= \langle f_1\left(-\hat{\bf p}_2\nabla+i\boldsymbol{\omega}_2{\bf L}+i\boldsymbol{\Omega}{\bf L}\right)f_2\rangle  + (1\leftrightarrow 2)\,.
\end{equation}
As before, we make the ansatz $f_i = \langle f_i\rangle +\delta{f}_i$ with the stationary solution $ \langle f_i\rangle$. Assuming that the spatial dependence of $\delta f_i$ is small compared to $\langle f_i\rangle$ we can replace the gradient term in Eq.~(\ref{eq:CLevolution}) via $\langle f_1\hat{\bf p}_2\nabla f_2\rangle \simeq-3/(4\pi)^2\hat{\bf p}_1\nabla{n}\hat{\bf p}_2{\bf K}\nabla{n}$. On the other hand, the second and third terms in Eq.~(\ref{eq:CLevolution}) are now non-trivial terms in the evolution that will lead to mixing and damping of multipoles. Analogous to the BGK-type approximation~\citep{Bhatnagar:1954zz} of Eq.~(\ref{eq:BGK}), we make the ansatz,
\begin{multline}\label{eq:BGK2}
\langle (i\boldsymbol{\omega}_1{\bf L}_1+i\boldsymbol{\omega}_2{\bf L}_2)f_1 f_2\rangle\\ \simeq -\left[\nu_{\rm r}(x)\frac{{\bf L}_1^2+{\bf L}_2^2}{2} +\nu_{\rm c}(x){\bf J}^2\right]\langle f_1f_2\rangle\,,
\end{multline}
with $x=\hat{\bf p}_1  \hat{\bf p}_2$ and ${\bf J}={\bf L}_1+{\bf L}_2$ . The relative and correlated scattering rates, $\nu_{\rm r}$ and $\nu_{\rm c}$, respectively, depend on the relative distance of the trajectories at early times, which in turn depends on the trajectories' opening angle. Note that Eq.~(\ref{eq:BGK2}) is necessarily symmetric under the interchange of ${\bf p}_1 \leftrightarrow {\bf p}_2$ as well as ${\bf p}_i\leftrightarrow -{\bf p}_i$ and is the most general linear approximation of all possible scalar combinations of ${\bf L}_1$ and ${\bf L}_2$. Note that as in the original BGK ansatz, we do not account for the possibility of anisotropic scattering in turbulent fields.

We now look for the power spectrum of a stationary solution of Eq.~(\ref{eq:CLevolution}) using the approximation (\ref{eq:BGK2}). The product $\langle f_1f_2\rangle$ can be expanded into eigenstates of ${\bf J}^2$ and ${J}_z$ and the power spectrum integral (\ref{eq:Celldef}) corresponds to projections onto the state $J=0$ with $M=0$. The rotation term corresponding to the regular magnetic field in Eq.~(\ref{eq:CLevolution}) does not contribute. Since the relaxation rates $\nu_{\rm r}(x)$ and $\nu_{\rm c}(x)$ in the approximation (\ref{eq:BGK2}) are assumed to depend only on $x=\hat{\bf p}_1\hat{\bf p}_2$ the relative rotation proportional to $({\bf L}^2_1+{\bf L}^2_2)/2 = {\bf L}^2$ is the only non-vanishing term.

For identical momenta and rigidities, the relative diffusion should vanish completely. One way to see this is by considering a toy model where the initial configuration of CRs is homogeneous~\citep{Ahlers:2013ima}. From this setup one can infer that the PSD at any time obeys $\partial_t \int {\rm d}\hat{\bf p} \langle f^2(t,{\bf r},{\bf p})\rangle = 0$. Therefore, in this case we must also have $\int {\rm d}\hat{\bf p}\nu_{\rm r}(1){\bf L}^2\langle f^2(t,{\bf r},{\bf p})\rangle = 0$. Since this is independent of the initial configuration we must therefore conclude that $\nu_{\rm r}(1)=0$.

A functional dependence can be obtained from a geometrical argument: two particles emitted locally under a small opening angle will separate with a reduced relative velocity $\Delta v\sim \sqrt{2(1-x)}$ as long as both particles are experiencing a strongly correlated magnetic field. This leads to the ansatz $\nu_{\rm r}(x) \propto (1-x)^p$ with $p = 1/2$. To allow for a weaker or stronger $x$-dependence, we have generalized this form to $\nu_{\rm r}(x) \propto (1-x)^p$, with $0<p<1$.

Under these assumptions and approximations, the stationary average $C_\ell$ spectrum then obeys the set of equations (see Appendix~\ref{app1})
\begin{equation}\label{eq:stationary}
\mathcal{Q}_1\delta_{\ell1} = \sum_k\langle C_k\rangle k(k+1)\frac{2k+1}{2}\int {\rm d}x\nu_{\rm r}(x)P_\ell(x)P_k(x)\,,
\end{equation}
with the effective dipole source term $\mathcal{Q}_1= {K}^{\rm s}_{ij}{\partial_i n\partial_j n}/(6\pi)$. Since the rhs of the previous equation is an expansion in terms of Legendre polynomials, we can easily derive the stationary spectrum as
\begin{equation}\label{eq:finalCl}
\langle C_\ell\rangle = \frac{3}{2}\frac{\mathcal{Q}_1}{\ell(\ell+1)}\int\limits_{-1}^1{\rm d}x\frac{x\,P_\ell(x)}{\nu_{\rm r}(x)}\,.
\end{equation}
With the assumed $\nu_{\rm r}(x) \propto (1-x)^p$, this simplifies to
\begin{equation}
\langle C_\ell\rangle = \frac{3 \mathcal{Q}_1}{2^{p}} \left( \frac{p(p-1)}{\ell(\ell+1)} + 1 \right) \frac{\Gamma(1-p) \Gamma(\ell+p-1)}{\Gamma(p) \Gamma(\ell-p+3)}\,.
\end{equation}
In the limit $\ell \gg 1$, $\Gamma(\ell+p-1) / \Gamma(\ell-p+3) \to \ell^{2p-4}$, and so $\langle C_\ell\rangle \propto \ell^{2p-4}$. For $p=1/2$, this is close to the $\ell^{-3}$ behavior exhibited by the data (cf.~Fig.~\ref{fig2}). 
Figure~\ref{fig2} also shows examples of the power spectrum (\ref{eq:finalCl}) for three models $\nu_r(x)\propto(1-x)^p$ with $p=1/3$, $1/2$, and $2/3$. While our ansatz $\nu_{\rm r}(x) \propto (1-x)^p$ is phenomenological in nature, a microscopic derivation of $\nu_{\rm r}(x)$, in quasi-linear theory, will be presented in future work.

\section{Conclusions}
\label{sec:conclusion}

We have discussed the appearance of small-scale anisotropy from the propagation of CRs in local turbulent magnetic fields. We have shown that this effect can be understood as a phenomenon of relative diffusion, where CRs with similar arrival directions experience correlated motion in the local magnetic environment. Our numerical simulation showed that the ensemble-averaged dipoles can be significantly larger than those expected from standard diffusion theory. In addition, the formalism predicts significant power in the ensemble-averaged small-scale multipoles. 

We have also provided an analytic framework that can describe the multipole power spectrum via an effective relative relaxation rate of multipole components in the evolution of the ensemble-averaged PSD products. This formalism is a natural extension of the well-known BGK approach. Assuming that the relative relaxation rate has a simple power-law dependence on the relative opening angle of CR arrival directions, we arrive at a power-law spectrum very similar to that obtained from simulations and the observed CR data.

{\it Acknowledgements.}---The authors would like to thank Paolo Desiati, Luke Drury, Gwenael Giacinti, Alex Lazarian, Vanessa L\'opez-Barquero, Martin Pohl, and Stefan Westerhoff for comments. We thank Dan Fiorino and Marcos Santander for discussions and providing the data points of Fig.~\ref{fig2}. M.A.~is supported by the National Science Foundation under grants OPP-0236449 and PHY-0236449. P.M.~is supported by DoE contract DE-AC02-76SF00515 and a KIPAC Kavli Fellowship.

\begin{appendix}

\section{Derivation of the Stationary Power Spectrum}\label{app1}

The decomposition of the PSD $f({\bf p})$ into spherical harmonics $Y_{\ell m}(\hat{\bf p})$ corresponds to a projection onto eigenfunctions $|\ell m\rangle$ of the angular momentum operator ${\bf L}$. The products of PSDs that appear in the definition of the power spectrum in Eq.~(\ref{eq:Celldef}) are then eigenfunctions of $|\ell_1 m_1\ell_2 m_2\rangle$ that can be decomposed into eigenfunctions $|JM\ell_1\ell_2\rangle$ of the total angular momentum operator ${\bf J}\equiv {\bf L}_1 +{\bf L}_2$. As noted earlier, the integral over arrival directions in Eq.~(\ref{eq:Celldef}) corresponds to a projection of $f_1f_2$ onto states with $J=0$ and $M=0$ and $\ell_1=\ell_2=\ell$, which can be seen from the expansion into spherical harmonics
\begin{equation}
|00\ell\ell\rangle  = \frac{(-1)^\ell}{\sqrt{2\ell+1}}\sum_mY^*_{\ell m}(\hat{\bf p}_1)Y_{\ell m}(\hat{\bf p}_2) = (-1)^\ell\frac{\sqrt{2\ell+1}}{4\pi}P_\ell(\hat{\bf p}_1\hat{\bf p}_2)\,,
\end{equation}
following from the Clebsch-Gordan coefficients
\begin{equation}
\langle00|\ell_1 m_1 \ell_2 -m_2\rangle = \delta_{\ell_1\ell_2}\delta_{m_1m_2}\frac{(-1)^{\ell_1-m_1}}{\sqrt{2\ell_1+1}}\,.
\end{equation}
Now, the BGK ansatz (\ref{eq:BGK2}) assumes that the scattering rates $\nu_r(x)$ and $\nu_c(x)$ are only functions of $x = \hat{\bf p}_1\hat{\bf p}_2$ and can thus be expanded into a sum over states with $J=0$ and $M=0$. The projection of Eq.~(\ref{eq:BGK2}) onto $|00\ell\ell\rangle$ can thus only receive contributions from the projection of the product of PSDs onto $J=0$ and $M=0$ states,
\begin{align*}
f_1f_2 = \sum_{pq}\sum_{rs}a_{pq}a_{rs}Y_{pq}(\hat{\bf p}_1)Y_{rs}(\hat{\bf p}_2)\to\sum_{pq}\sum_{rs}a_{pq}a_{rs}\langle00| pqrs\rangle |00\rangle=\frac{1}{4\pi}\sum_p(2p+1)P_p(\hat{\bf p}_1\hat{\bf p}_2)C_p\,.
\end{align*}
Finally, the stationary solution (\ref{eq:stationary}) has the form $Q_1\delta_{\ell1} = \int {\rm d}x P_\ell(x) K(x)$ and hence $K(x) \propto P_1(x) = x$. From this we immediately arrive at Eq.~(\ref{eq:finalCl}).

\end{appendix}

\end{document}